\documentclass[sigconf]{acmart}

\usepackage{amsmath,amsfonts}
\usepackage{algorithmic}
\usepackage{textcomp}
\usepackage{xcolor}
\usepackage{booktabs}
\usepackage{array}
\usepackage{multirow}
\usepackage{multicol}
\usepackage{enumitem}
\usepackage{subfig}
\usepackage{bm}
 
\usepackage{amssymb}

\newcommand{\Mat}[1]{\bm{#1}}
\newcommand{\Vector}[1]{\bm{#1}}
\newcommand{\Set}[1]{\mathcal{#1}}
\newcommand{\Real}{\mathbb{R}}
\newcommand{\Loss}{\mathcal{L}}
\newcommand{\Normal}[1]{\mathrm{#1}}

\AtBeginDocument{%
  \providecommand\BibTeX{{%
    \normalfont B\kern-0.5em{\scshape i\kern-0.25em b}\kern-0.8em\TeX}}}

\copyrightyear{2022}
\acmYear{2022}
\setcopyright{acmcopyright}
\acmConference[WSDM '22]{Proceedings of the 15th ACM International Conference on Web Search and Data Mining}{February 21--25, 2022}{Phoenix, AZ, USA}
\acmBooktitle{Proceedings of the 15th ACM International Conference on Web Search and Data Mining (WSDM '22), February 21--25, 2022, Phoenix, AZ, USA}
\acmPrice{15.00}
\acmDOI{10.1145/1122445.1122456}
\acmISBN{978-x-xxxx-xxxx-x/YY/MM}

\begin{document}

\title{A Practical Two-stage Ranking Framework for Cross-market Recommendation}

\author{Zeyuan Chen$^{1}$, He Wang$^{2}$, Xiangyu Zhu$^{3}$, Haiyan Wu$^{1}$, Congcong Gu$^{4}$, Shumeng Liu$^{2}$, Jinchao Huang$^{2}$, Wei Zhang$^{1*}$}
\thanks{*All the corresponding to Wei Zhang}
\affiliation{%
  \institution{$^{1}$East China Normal University, $^{2}$Xiaomi Inc., $^{3}$JD.com, $^{4}$Pingan Inc.}
  \country{}
}
\email{chenzyfm@outlook.com, {wanghe11,liushumeng,huangjinchao1}@xiaomi.com, zhuxiangyu3@jd.com}
\email{gucongcong169@pingan.com.cn, hywuu@outlook.com, zhangwei.thu2011@gmail.com}

\renewcommand{\shortauthors}{Chen and Wang, et al.}

\begin{abstract}
Cross-market recommendation aims to recommend products to users in a resource-scarce target market by leveraging user behaviors from similar rich-resource markets, which is crucial for E-commerce companies but receives less research attention. In this paper, we present our detailed solution adopted in the cross-market recommendation contest, i.e., WSDM CUP 2022\footnote{\url{https://xmrec.github.io/wsdmcup/}}. To better utilize collaborative signals and similarities between target and source markets, we carefully consider multiple features as well as stacking learning models consisting of deep graph recommendation models (Graph Neural Network, DeepWalk, etc.) and traditional recommendation models (ItemCF, UserCF, Swing, etc.). Furthermore, We adopt tree-based ensembling methods, e.g., LightGBM, which show superior performance in prediction task to generate final results. We conduct comprehensive experiments on the XMRec dataset, verifying the effectiveness of our model. The proposed solution of our team \textit{WSDM\_Coggle\_} is selected as the second place submission\footnote{The source code is available at \url{https://github.com/loserChen/WSDM_CUP_Rec_2022}}.
\end{abstract}



\maketitle

\section{Introduction}\label{sec:intro}
Recommender systems (RS) are ubiquitous in online platforms and mobile applications, such as e-commerce. 
Cross-domain recommendation is one kind of RS that leverages the interactions of overlapping items in source domains to benefit the recommendations in a target domain. 
There has been a number of studies with various domain definitions and recommendation scenarios~\cite{yuan2019darec,hu2018conet}. However, few studies have been conducted for cross-market recommendation.

The recommendation scenario of cross-market recommendation~\cite{bonab2021cross} is that the model learns from interactions of overlapping items in different markets to improve recommendation performance in a target domain, hoping to utilize information from rich source markets. 
Although it is crucial for E-commerce companies to combine different markets of various countries to solve the cold-start and data sparsity problems~\cite{ZhangW15} occurred in a resource-scarce target market, little progress is made partly due to the lack of publicly available experimental data. 

Thanks to the cross-market recommendation contest in WSDM CUP 2022, it provides the XMRec dataset which contains abundant user-item interaction records in different markets for further research purpose. To address this challenge, we introduce a practical two-stage ranking framework which contains hybrid model ranking and GBDT learning~\cite{wang2017combining}.

\begin{figure*}[!t]
    \centering
	\includegraphics[width=.7\linewidth]{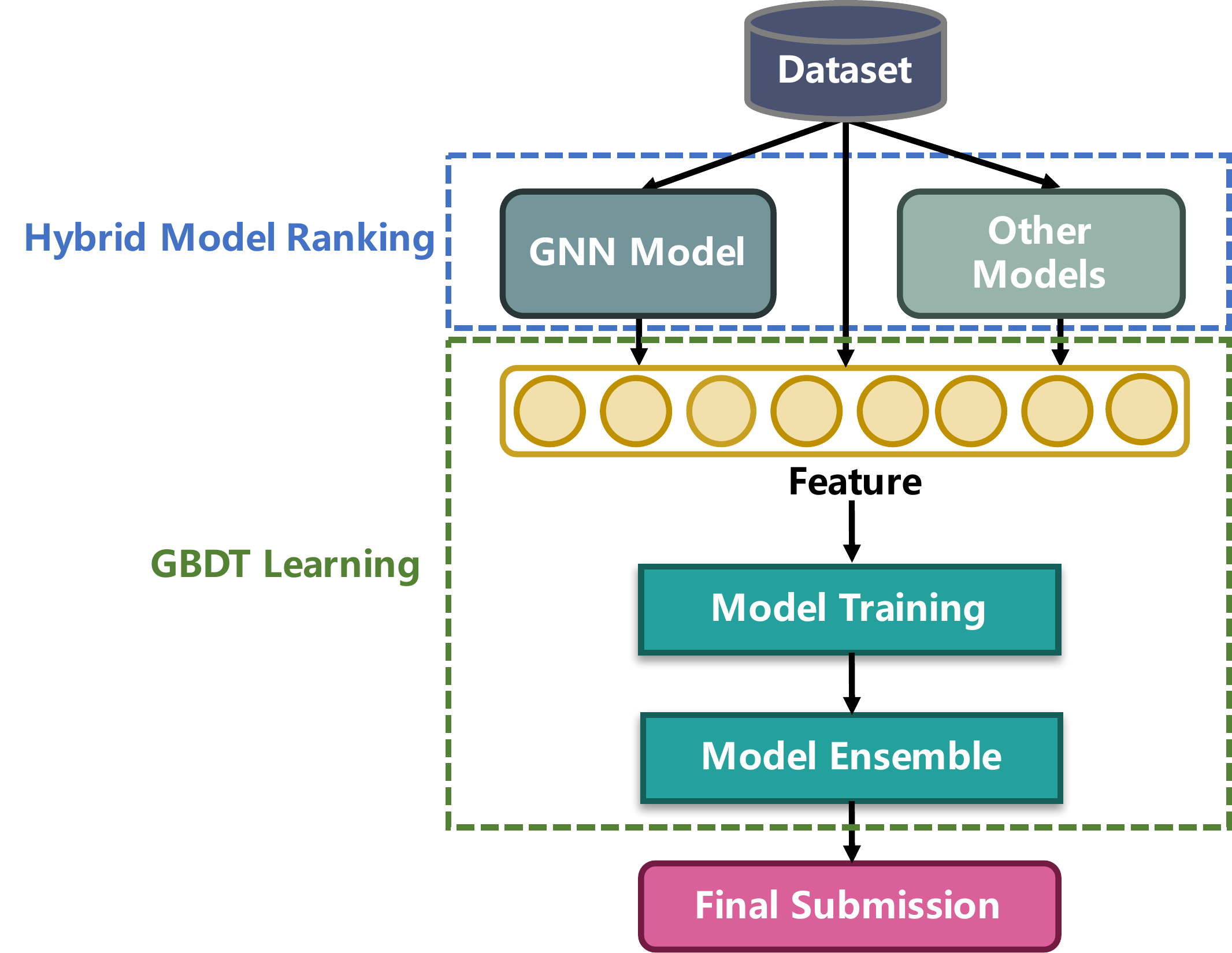}
    \caption{The whole framework of our model to address the challenge of cross-market recommendation .}
    \label{fig:model}
\end{figure*}

The rest of the paper is organized as follows.
We first provide a brief data analysis in Section~\ref{sec:dataset}. Section~\ref{sec:ranking} is about hybrid models ranking. GBDT learning is in Section~\ref{sec:gbdt}, which includes feature engineering, models building, and ensemble modeling. Experimental results are illustrated in Section~\ref{sec:exp}. Finally, we make a conclusion in Section~\ref{sec:con}. The overall framework of our approach is shown in Figure~\ref{fig:model}.

\section{Dataset}\label{sec:dataset}
The dataset\footnote{\url{https://github.com/hamedrab/wsdm22_cup_xmrec}} provided by the sponsor contains five folders: s1, s2, s3,
t1, and t2. The folders s1, s2, and s3 contain the data of the source markets (train.tsv, train\_5core.tsv, valid\_qrel.tsv, and valid\_run.tsv) for training and validating. The other folders t1 and t2 involve the data of the target market. Inside each, there are the training set (train.tsv and train\_5core.tsv) and the public/private test set (valid\_qrel.tsv and valid\_
run.tsv/test\_run.tsv), respectively. For ease of use, we simply concatenate train.tsv and train\_5core.tsv by deleting the repeated samples and generate train\_merge.tsv for each folder. Concretely, train\_merge.tsv contains the training data with the fields of userId, itemId, and rating.
valid\_qrel.tsv is the validation positive samples, with a data structure similar to train.tsv.
valid\_run.tsv is the validation samples wherein each row has 99 negative samples and 1 positive sample for each unique userId.
test\_run.tsv is the test candidate samples with the same positive or negative sample ratio as valid\_run.tsv. The basic statistics of the datasets are summarized in Table~\ref{tbl:stat}.

\begin{table}[!t]
\centering
\caption{Statistics of the datasets.}\label{tbl:stat}
\begin{tabular}{cccc}
\hline
\textbf{Dataset}  & \# Users  & \# Items  & \# Interactions \\ \hline
 s1 & 77,776 & 11,807 & 793,300 \\
 s2 & 20,311 & 3,408 & 794,477 \\
 s3 & 8,568 & 2,332 & 379,092 \\
 \hline
 t1 & 9,955 & 3,559 & 599,600 \\
 t2 & 18,504 & 8,941 & 1,216,378 \\
\hline
\end{tabular}
\end{table}

\section{Hybrid Model Ranking}\label{sec:ranking}
In order to improve the diversity of the model, we choose a few deep graph recommendation models and traditional recommendation models to implement result prediction.
Since we do not significantly change these models, we just simply introduce them used in the contest.
In what follows, we mainly introduce our ranking-augmented graph neural network based on practical findings during this contest.

\subsection{Ranking-augmented Graph Neural Network}
Before delving into the computational formulas of this module, we need to firstly clarify how to use the dataset for result prediction. We use the training set to train with sampling 99 negative items for each positive interaction, public test set to validate model performance and test candidate samples to predict without using any data from source markets, due to no distinct performance improvement by trying cross-market information. In order to distinguish the degree to which a user prefers an item, we build user-item bipartite graph $\Set{G}$ based on explicit feedback, i.e., rating. As such, we form the following formula, i.e., $((rating)/10+0.5)*0.95+0.05/2$, which decides the fine-grained edge weights in graph $\Set{G}$. Assume there are $M$ users and $N$ items occurring in the target markets, then we have a node feature matrix $\Mat{X}\in\Real^{(M+N)\times d}$ and an adjacency matrix $\Mat{A}^{(M+N)\times(M+N)}$ for the graph.

By convention, GNN performs representation along with the edges of graph $\Set{G}$, which is defined as follows:
\begin{equation}\label{eq:propagation}
    \Mat{X}_{l+1} = \Mat{\hat{A}}\Mat{X}_{l}+\Mat{X}_{l}\odot \Mat{\hat{A}}\Mat{X}_{l}\,,
\end{equation}

where $\Mat{\hat{A}}=\Mat{D}^{-0.5}\Mat{A}\Mat{D}^{-0.5}$ denotes the normalized adjacency matrix without self loops and $l$ is the index of the propagation layer.
We let $\Mat{X}_{0}=\Mat{X}$, consisting of input user and item representations. Distinct from conventional graph convolution networks which do not consider information transfer between the adjacent layer, we additionally encode the similarity between $\Mat{X}_{l}$ and $\Mat{X}_{l+1}$ so that more messages from similar nodes can be passed~\cite{zhang2021learning}. By doing this, it can boost the ranking performance based on our extensive experiments.

After propagation of $L$ layer, we obtain multiple layer-wise representations, namely \{$\Mat{X}_0;\Mat{X}_1;\dots;\Mat{X}_L$\}. Since the representations from different layers emphasize different semantics, we simply perform average-pooling on them to constitute  embedding $\Mat{\bar{X}}$. In order to mitigate the popularity bias in recommendation, we adopt a simple but effective solution, i.e., the $L_2$ based normalization operation, on $\Mat{\bar{X}}$ so as to obtain final embedding $\Mat{\tilde{X}}$.

Based on the final embedding $\Mat{\tilde{X}}$, we could have two representations for user $u$ and item $i$, i.e., $\Vector{e}_u$ and $\Vector{e}_i$. To predict the potential interaction probability for the considered user-item pair, we proceed some changes to fuse GMF and MLP referred to classical neural collaborative filtering method NeuMF~\cite{HeLZNHC17} and improve the ability of model prediction, which can be defined as follows:
\begin{equation}
    \hat{y}_{ui}=\sigma(\Vector{z}_{GMF}+\Vector{z}_{MLP})=\sigma(\Vector{h}^T(\Vector{e}_u\odot\Vector{e}_i)+\Normal{MLPs}(\Vector{e}_u\oplus\Vector{e}_i))\,,
\end{equation}
where $\Vector{h}\in \Real^{d\times1}$ is a trainable weight vector. ReLU is adopted as the middle-layered activation function in $\Normal{MLPs}$. $\odot$ denotes the element-wise product of vectors and $\oplus$ means the concatenation operation. $\sigma$ is the sigmoid function.

For training this module, we employ the cross-entropy loss, which is given by:
\begin{equation}\label{eq:loss}
    \Loss = -\big(y_{ui}\log\hat{y}_{ui}+(1-y_{ui})\log(1-\hat{y}_{ui})\big)\,.
\end{equation}
To avoid overfitting user preferences, the label smoothing trick~\cite{yuan2020revisiting} is adopted based on corresponding ratings. Finally, we keep the prediction score of public/private test set from this module for the follow-up experiments.

\subsection{Other Used Recommendation Models}
Unlike the way of using data mentioned above, the recommendation models adopt another way to utilize the datasets. Here we use the training set to build the corresponding features and give predictions on the public/private test set. Similarly, we keep the corresponding score for further research. It is worth noting that the training set used here includes all the data from source markets and target markets so as to learn cross-market information and improve model performance. 
Unless otherwise specified, the data usage setting is used.
The used recommendation models will be introduced simply. Generally, these models are usually applied in the recall stage in industrial scenarios.

\subsubsection{ItemCF}
The core idea behind ItemCF~\cite{sarwar2001item} is to recommend items that is similar to items of interest to the user in the past. ItemCF focuses on maintaining users' historical interests to make recommendations more personalized and reflects users' own interest inheritance.

\subsubsection{UserCF}
Similarly, UserCF tends to recommend items that are liked by other users with similar interests. And UserCF focuses on the hot spots of small groups similar to users' interests. The recommendation results are more social and reflect the popularity of items in users' interest groups.

\subsubsection{Swing}
Swing considers local graph structure relations such as user-item-user. For users who click on items i and j together, the fewer items they click on, the more similar items i and j are.

\subsubsection{PersonalRank}
PersonalRank~\cite{yang2018nearest} is a graph-based algorithm which utilizes random walk iteratively and the access probability of nodes gradually converges.

\section{GBDT Learning}\label{sec:gbdt}
Thanks to high efficiency and the superior ability of GBDT, we choose GBDT as our base model which is a popular machine learning algorithm, and has quite a few effective implementations such as XGBoost~\cite{chen2016xgboost}, LightGBM~\cite{ke2017lightgbm}, and CatBoost~\cite{dorogush2018catboost}. By convention, feature engineering is a pivotal process especially for these tree-based ensembling methods. In order to improve the fitting ability of the whole framework, we investigate novel learning strategies to get different models. Finally, ensembling modeling is adopted which always works very well in different contests.

\subsection{Feature Engineering}
Except for regarding the prediction results generated by the above recommendation models as ranking features, we also construct statistical features, embedding features, and distance features.
These features could capture useful information across different markets, so as to benefit the contest.

\subsubsection{Statistical Features}
Taking user statistical features as an example, we generate the corresponding features based on the userId, itemId and rating fields such as the count of purchasing, the number of types of items purchased as well as computing minimum, median, maximum, mean and standard deviation of ratings given by each user. And it is analogous to item statistical features.

\subsubsection{Embedding Features}
We build user and item embedding features respectively and the following introduces the embeddings we used in this contest:
\begin{itemize}[leftmargin=*]
\item TF-IDF~\cite{jing2002improved} is always applied to NLP and information retrieval fields, which is a statistical quantity for measuring the importance of a word with respect to a document. As for recommendation, we can think of users and items as words to achieve a similar goal. To avoid dimension explosion, the SVD method~\cite{aharon2006k} is used to remove unimportant components.

\item Word2Vec~\cite{goldberg2014word2vec} is a neural network model to generate vector representations of words, which also can be used for generating user and item representations.

\item DeepWalk~\cite{perozzi2014deepwalk} could learn network embedding by proceeding truncated random walk based on a user-item interaction graph.

\end{itemize}

\subsubsection{Distance Features}
Based on the embedding features we have constructed, we compute cosine distance, Manhattan distance, Jaccard coefficient, euclidean distance and Pearson correlation coefficient to measure the correlation between the two embeddings of the same type from users and items, and the correlation
values are used as our distance features.

\subsection{Model Training}
Here we introduce learning strategies in detail. Based on features introduced above, we use multi-fold cross validation against the public test set of target markets to finish predicting test candidate samples.

In essence, recommendation is the task of learning to rank, which mainly includes three types of learning strategies, i.e., pointwise, pairwise and listwise. The pointwise learning regards the ranking problem as a classification or regression problem. However, the pairwise learning does not care about the specific value but only considers the relative order. As for the listwise learning, it tackles the ranking problem directly by optimizing the defined loss function on a list of items. 

Thanks to the effective implementation of GBDT's pointwise and pairwise learning versions. We use LightGBM, XGBoost, CatBoost to train by pointwise learning and LGBMRanker~\cite{jankiewicz2019boosting} to train by pairwise learning. Thus, we could obtain four final models.

\subsection{Ensemble Modeling}
In the model ensemble stage, we simply adopt the weighted average operation to fuse these four models to get the final results. In fact, we have stacked predictions from multiple models to build these tree-based ensembling models, which also belongs to the model ensemble operation.

\section{Experiments}\label{sec:exp}
We evaluate our method on the XMRec dataset provided by the contest. Table~\ref{tbl:performance-comp} presents the overall performance of our model and all the adopted baselines, from which we have the following key observations:
\begin{itemize}
    \item In the first part of the table, GNN achieves the best performance on both datasets compared to other traditional models. It may be attributed to the superior power of GNN and our effective module design.
    \item LightGBM* and LGBMRanker* are the models without using ranking features generated by the hybrid ranking models. By fine-grained feature engineering, they could also generate comparable performance.
    \item As for the third part of the table, it proves that the GBDT models could obtain major improvements by stacking learning models and multiple models ensemble.
\end{itemize}

\begin{table}[!t]
\centering
\caption{Main results w.r.t. NDCG@10 and HR@10 for cross-market recommendation on target market datasets. The best and second-best performed methods in each metric are highlighted in “bold” and underline, respectively.}\label{tbl:performance-comp}
\resizebox{\linewidth}{!}{
\begin{tabular}{ccc|cc} 
\hline
\multirow{2}*{Method}& \multicolumn{2}{c}{t1}& \multicolumn{2}{c}{t2}\\\cline{2-5}
&\multicolumn{1}{c}{NDCG@10} &\multicolumn{1}{c}{HR@10} &\multicolumn{1}{c}{NDCG@10} &\multicolumn{1}{c}{HR@10} \\ \hline

GNN
&\multicolumn{1}{c}{0.7131} &\multicolumn{1}{c}{0.7968} &\multicolumn{1}{c}{0.6226} &\multicolumn{1}{c}{0.7255} \\

ItemCF
&\multicolumn{1}{c}{0.5955} &\multicolumn{1}{c}{0.6782} &\multicolumn{1}{c}{0.5235} &\multicolumn{1}{c}{0.6136} \\

UserCF
&\multicolumn{1}{c}{0.6136} &\multicolumn{1}{c}{0.6948} &\multicolumn{1}{c}{0.5472} &\multicolumn{1}{c}{0.6547} \\

Swing
&\multicolumn{1}{c}{0.6049} &\multicolumn{1}{c}{0.6834} &\multicolumn{1}{c}{0.5489}
&\multicolumn{1}{c}{0.6350} \\

PersonalRank
&\multicolumn{1}{c}{0.5922} &\multicolumn{1}{c}{0.6756} &\multicolumn{1}{c}{0.5582}
&\multicolumn{1}{c}{0.6618} \\\hline

LightGBM*
&\multicolumn{1}{c}{0.7197} &\multicolumn{1}{c}{0.8224} &\multicolumn{1}{c}{0.6286}
&\multicolumn{1}{c}{0.7421} \\

LGBMRanker*
&\multicolumn{1}{c}{0.7039} &\multicolumn{1}{c}{0.7801} &\multicolumn{1}{c}{0.6381}
&\multicolumn{1}{c}{0.7506} \\\hline

CATBoost
&\multicolumn{1}{c}{0.7310} &\multicolumn{1}{c}{0.8246} &\multicolumn{1}{c}{0.6371}
&\multicolumn{1}{c}{0.7474} \\

XGBoost
&\multicolumn{1}{c}{0.7354} &\multicolumn{1}{c}{0.8283} &\multicolumn{1}{c}{0.6391} &\multicolumn{1}{c}{0.7512} \\

LightGBM
&\multicolumn{1}{c}{0.7362} &\multicolumn{1}{c}{0.8324} &\multicolumn{1}{c}{\underline{0.6399}} &\multicolumn{1}{c}{0.7534} \\

LGBMRanker
&\multicolumn{1}{c}{\underline{0.7388}} &\multicolumn{1}{c}{\underline{0.8356}} &\multicolumn{1}{c}{0.6388} &\multicolumn{1}{c}{\underline{0.7555}} \\\hline

\textbf{Model Ensemble}
&\multicolumn{1}{c}{\textbf{0.7393}} &\multicolumn{1}{c}{\textbf{0.8369}} &\multicolumn{1}{c}{\textbf{0.6457}} 
&\multicolumn{1}{c}{\textbf{0.7610}} \\
\hline
\end{tabular}
}
\end{table}

\section{Conclusion}\label{sec:con}
In this paper, we have introduced our practical two-stage ranking framework for the cross-market recommendation competition of the WSDM Cup 2022. Our team ranks the second place on the final leaderboard\footnote{\url{https://competitions.codalab.org/competitions/36050\#results}} with an excellent performance very close to the first place. In our solution, we first conduct various models to give predictions. After that, we train 4 GBDT models by utilizing different learning strategies and implementations of GBDT. Finally, we ensemble these 4 models by weighted average operation. The comprehensive experiment results demonstrate the superiority and effectiveness of our model.

\begin{acks}
We thank everyone associated with organizing and sponsoring the
WSDM Cup 2022. 
\end{acks}

\bibliographystyle{ACM-Reference-Format}
\bibliography{references}

\end{document}